# Reduced dielectric screening and enhanced energy transfer interactions in single and few-layer MoS$_2$

Ferry Prins[1,*], Aaron J. Goodman[2], and William A. Tisdale[1,*]

[1] Department of Chemical Engineering, Massachusetts Institute of Technology, 77 Massachusetts Avenue, Cambridge, MA 02139, USA.

[2] Department of Chemistry, Massachusetts Institute of Technology, 77 Massachusetts Avenue, Cambridge, MA 02139, USA.

tisdale@mit.edu and prins@mit.edu

We report highly efficient non-radiative energy transfer from cadmium selenide (CdSe) quantum dots to monolayer and few-layer molybdenum disulfide (MoS$_2$). The quenching of the donor quantum dot photoluminescence increases as the MoS$_2$ flake thickness decreases, with the highest efficiency (>95%) observed for monolayer MoS$_2$. This counterintuitive result arises from reduced dielectric screening in thin layer semiconductors having unusually large permittivity and a strong in-plane transition dipole moment, as found in MoS$_2$. Excitonic energy transfer between a 0D emitter and a 2D absorber is fundamentally interesting and enables a wide range of applications including broadband optical down-conversion, optical detection, photovoltaic sensitization, and color shifting in light-emitting devices.

The recently isolated monolayers of transition metal dichalcogenides (TMDCs) are a promising new class of materials for applications in optoelectronic devices, combining high charge-carrier mobility with a direct optical bandgap.[1–3] The strongly anisotropic shape of these atomically thin semiconductors confines charge carriers to the two-dimensional plane of the material,[4] resulting in enhanced Coulomb interactions as well as strong resonant excitonic absorption features.[5] These unique properties have already led to the use of TMDCs, such as $MoS_2$, $MoSe_2$ and $WS_2$, in a variety of optoelectronic devices, including photodetectors,[6–8] solar convertors,[9–11] and light emitting diodes.[11,12]

The excitonic nature of optical excitations in monolayer TMDCs opens up possibilities to construct hybrid architectures in which different nanomaterials are coupled through efficient non-radiative energy transfer. Energy transfer between donor and acceptor materials can be used to enhance optoelectronic device performance, enabling broadband optical down-conversion[13] and color shifting in light-emitting devices[14,15] as well as energy transfer sensitized photovoltaics.[16,17] One possible approach is to interface two-dimensional TMDCs with zero-dimensional semiconductor nanocrystals, also known as quantum dots (QDs). QDs profit from size-tunable and broadband absorption,[18] narrow emission spectra, high photoluminescence quantum yield and can in some cases accommodate multi-exciton generation.[19–21] In a hybrid energy transfer architecture, QD donors can enhance the absorptive properties of TMDC optoelectronic devices, yielding more efficient solar cells and photodetectors. Alternatively, the high mobility TMDC transistors can provide efficient donation of electrically generated excitons to the QDs and provide tunable light emission.

Here, we demonstrate non-radiative energy transfer from colloidal quantum dots to exfoliated monolayer and few-layer molybdenum disulfide (see Figure 1a). We use high quantum yield (>80%) CdSe/CdZnS core/shell quantum dots having an emission spectrum that overlaps with the strong $MoS_2$ direct excitonic absorption features[1] (see Figure 1b). We observe a counterintuitive phenomenon wherein the energy transfer rate (and efficiency) monotonically decreases as the number of $MoS_2$ layers increases – despite unchanging $MoS_2$ absorption in this region of the spectrum and the expectation that additional $MoS_2$ layers should offer additional pathways for energy transfer.[22] These observations are explained in the context of recent theoretical predictions[23,24] of anomalous dielectric screening effects that can present themselves in thin semiconductor structures.

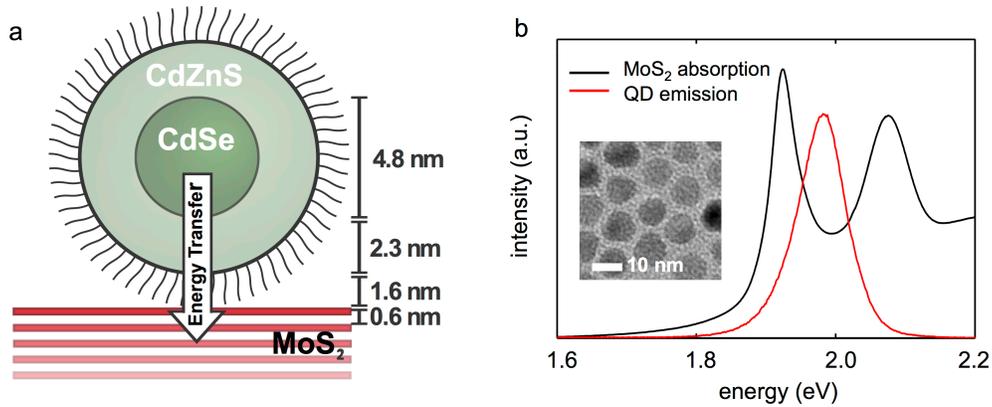

**Figure 1 (a)** Schematic of the hybrid QD/$MoS_2$ energy transfer system with indicated length scales. **(b)** Monolayer $MoS_2$ absorption spectrum (obtained from ref. [25]), overlaid with the film-based CdSe/CdZnS emission spectrum. Inset: TEM of a monolayer of the CdSe/CdZnS QDs used in this study.

Monolayer and few-layer MoS$_2$ flakes were deposited on Si/SiO$_2$ substrates by mechanical exfoliation from bulk single crystalline MoS$_2$ (Graphene Supermarket). Core-shell CdSe/CdZnS quantum dots (QD Vision Inc.) were deposited onto the samples using spin-coating from a dilute solution (0.5 mg/mL in toluene, 1500 rpm) to form a spatially homogeneous QD sub-monolayer that is ≤1 QD thick in all locations. The CdSe QD cores are 4.8 nm in diameter and surrounded by a 2.3 nm thick Cd$_{0.5}$Zn$_{0.5}$S shell that is capped with a layer of organic ligands (octadecyl phosphonic acid) 1.6 nm in length (see Figure 1b for a TEM image). The quantum dots are well passivated, resulting in solution phase quantum yields of > 80% and a native 1/e lifetime of 20.4 ± 0.1 ns.

Scanning photoluminescence lifetime measurements were performed using an inverted microscope (Nikon, Ti Eclipse), equipped with an x-y piezo stage (Nanonics, MV2000). The samples were excited using a 405 nm pulsed laser diode (LDH-D-C-405M, Picoquant, 10 MHz repetition rate, 0.5 ns pulse duration) with an average pulse fluence of 60 nJ / cm$^2$ across a near-diffraction limited spot ~1 μm in diameter (Nikon, 20× objective, N.A. = 0.4). The photoluminescence was collected with the same objective, passed through a dichroic mirror and long-pass filter, and focused onto a Si avalanche photodiode (Micro Photon Devices, PDM50, 32 ps resolution). The detector was connected to a counting board for time correlated single photon counting (Picoquant, PicoHarp 300). During laser scanning fluorescence lifetime imaging, the sample was scanned by the piezo stage, recording a fluorescence lifetime trace for each pixel. Raman imaging was performed using a scanning Raman microscope (Horiba, LabRAM) with a 532 nm light source (Kaiser Optical Systems, Inc., Invictus).

Figure 2a shows a bright-field optical micrograph of a MoS$_2$ flake prior to QD deposition. The different colors indicate several different thicknesses present in this sample.[26] To determine the layer thickness of each facet, we performed micro-Raman measurements of this same flake. The energy difference between the A$_{1g}$ and E$^1_{2g}$ Raman modes is sensitive to the layer thickness and can be used to quantify the thickness of few-layer MoS$_2$.[27] We perform this characterization for a variety of flakes and identify facets ranging from bulk-like thickness (>8 layers) down to monolayer thickness. The sample shown in Figure 2a contains facets ranging from three to more than eight layers thick.

A sub-monolayer of quantum dots was deposited uniformly across the different samples, and we used laser-scanning microscopy to construct maps of QD exciton lifetime (Figure 2b). Compared to the much larger volume-integrated absorption cross-section and efficient emission of the QDs, MoS$_2$ was only weakly fluorescent and contributed negligibly to total photoluminescence. For each pixel in the laser scanning microscopy image we recorded a fluorescence lifetime decay curve and extracted the 1/e lifetime. For all samples, we found that the fluorescence lifetime of QDs away from the flake was close to their native lifetime of 20.4 $\pm$ 0.1 ns. In contrast, the fluorescence lifetime of the QDs on top of the MoS$_2$ was significantly shortened, indicating strong quenching of the QD photoluminescence by energy transfer to MoS$_2$. Interestingly, we observed that the lifetime varied as a function of flake thickness, with strongest QD quenching occurring for the thinnest regions of MoS$_2$ (discussed below).

In Figure 2b we show the lifetime map of the multifaceted sample of figure 2a. The color scale is saturated above 5 ns to highlight the thickness dependence of the quenching. The bulk-like section of this sample has a spatially averaged lifetime of 2.82 ± 0.04 ns while the trilayer section has a spatially averaged lifetime of 1.70 ± 0.01 ns. We can express the lifetime variations in terms of an energy transfer efficiency, given by $\eta_{ET} = 1 - (\tau_{DA}/\tau_D)$, in which $\tau_D$ is the native 1/e lifetime of the donor QDs on SiO$_2$ (i.e. in the absence of MoS$_2$) and $\tau_{DA}$ the 1/e lifetime of the QDs on top of the MoS$_2$ acceptor. The resulting efficiency map is shown in Figure 2c. For all flakes that we investigated, we found that the energy transfer efficiency is consistently above 80%, reaching as high as 95.1 ± 0.1% for monolayer sections.

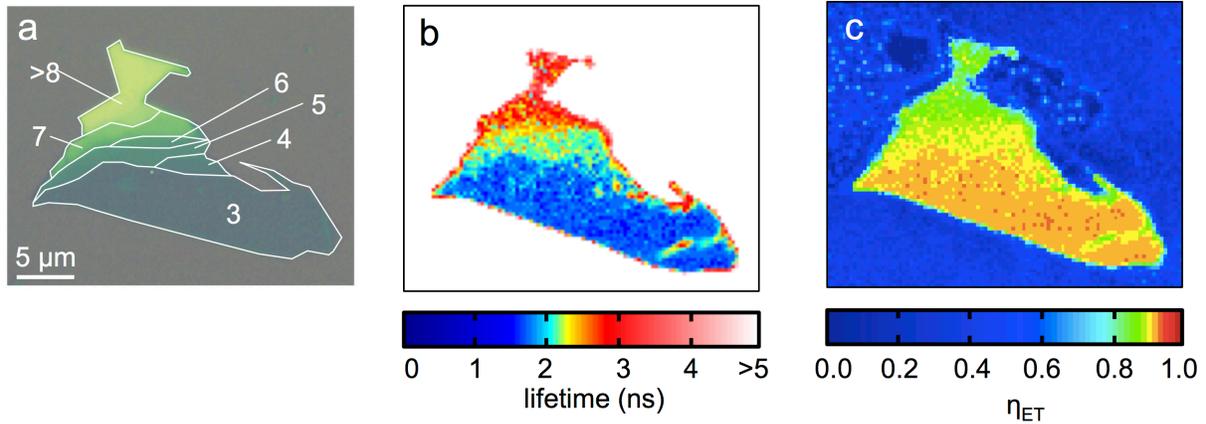

**Figure 2** (a) Optical micrograph of a mechanically exfoliated MoS$_2$ flake with indicated layer thicknesses of the different facets. (b) QD fluorescence lifetime map of the same flake with a saturated color scale to emphasize the lifetime variations within the flake area. (c) Energy transfer efficiency map of the same flake. Scale bar is identical for all three panels.

The observed quenching of QD fluorescence is consistent with non-radiative energy transfer from the QDs to MoS$_2$ through dipolar interactions.[28] It should be noted, though, that the quenched intensity and shortened lifetime of the CdSe/CdZnS QDs could in principle also originate from individual charge transfer to MoS$_2$. However, charge transfer would require tunneling across the ~ 4 nm thick barrier of the combined organic ligand and inorganic CdZnS shell, and is therefore unlikely to have a significant contribution to the quenching.[29] To further confirm non-radiative energy transfer as the responsible mechanism, we performed control experiments with InAs QDs having a bandgap energy smaller than that of $MoS_2$. Figure 3 shows representative fluorescence decay histograms of both the CdSe and InAs QDs. In contrast to the CdSe QDs and consistent with the energy transfer mechanism, the photoluminescence lifetime of the InAs QDs remains unchanged when placed on top of a $MoS_2$ surface.

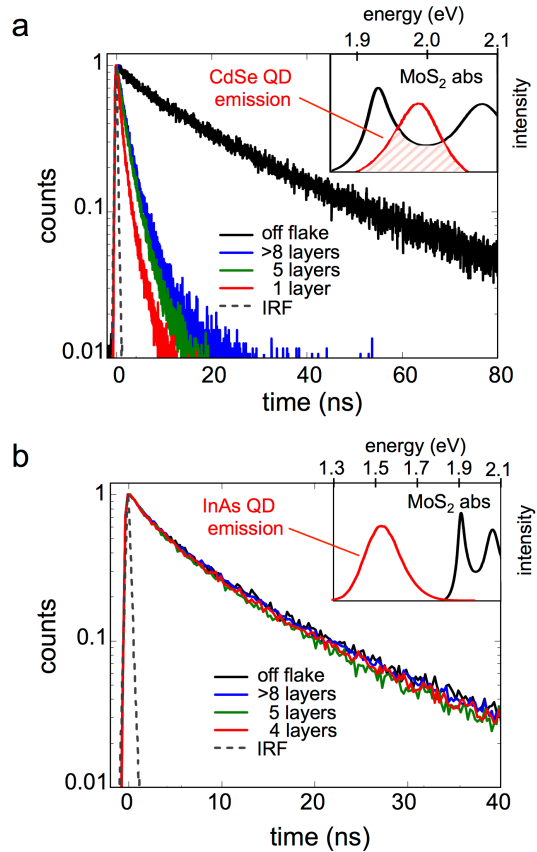

**Figure 3 (a)** Fluorescence lifetime histograms of CdSe QDs on different thicknesses of $MoS_2$. The black curve is obtained from QDs on top of $SiO_2$. The dashed gray curve represents the intrument's response function (IRF). Inset: spectral overlap between CdSe QD emission and $MoS_2$ absorption spectra. **(b)** Same as (a) but for InAs quantum dots with a bandgap energy below the $MoS_2$ bandgap.

We now turn to the layer thickness dependence of energy transfer. The resonant MoS$_2$ absorption features shown in Figure 1b are due to spin-orbit split direct excitonic transitions at the K-point of the MoS$_2$ Brillouin zone.[30] The oscillator strength and spectral shape of these transitions are largely unaffected by the MoS$_2$ flake thickness,[1,2,4,31] and therefore the large (>300%, see below) increase in the energy transfer rate with decreasing thickness cannot be explained by changes in the spectral overlap. Consequently, the observation of more efficient energy transfer for thinner MoS$_2$ acceptor layers is inconsistent with the most basic approximations of energy transfer, as described in Förster theory.[32] Within this model, the energy transfer rate is calculated assuming dipolar coupling between donor and acceptor point dipoles. Förster theory can be expanded to describe energy transfer between a point emitter and an acceptor surface, by treating the acceptor surface as an array of non-interacting point dipoles.[33] A larger volume of acceptor material automatically leads to more efficient energy transfer, because each additional dipole constructively contributes to the energy transfer rate. This theoretical framework has been successfully applied to describe energy transfer in a variety of nanomaterial systems, including energy transfer in colloidal quantum dot assemblies,[34,35] between emitters and graphene surfaces,[22,24] as well as from molecular dyes to plasmonic modes in thin metallic films.[36] Moreover, as expected within this framework but contrasting with our own observations, recent experiments using graphene as the acceptor for QD donors showed a monotonically *decreasing* energy transfer rate for decreasing thickness of few-layer graphene flakes.[22]

Interestingly, it has been suggested that the assumption of non-interacting point dipoles in Förster theory is an over-simplification in the limit of thin semiconductor acceptor films with high polarizability.[23,24] The thin film geometry can lead to enhanced energy transfer as a result of reduced dielectric screening of in-plane components of the donor dipole field.[37] This geometric effect is accentuated in acceptor media with in-plane oriented transition dipole moments,[23] and may in extreme cases lead to the counter-intuitive result where thinner acceptor layers yield more efficient energy transfer.[23] Thin $MoS_2$ flakes are characteristic examples of materials that should exhibit this effect. $MoS_2$ has a large permittivity,[25,26] and its dielectric function is reported to be highly anisotropic, with the resonant transitions polarized completely in the plane of the material.[38]

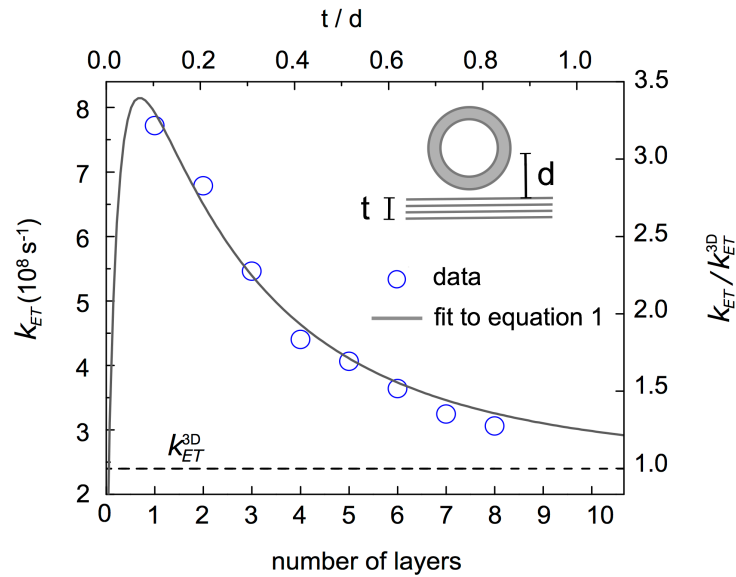

**Figure 4** Energy transfer rate as a function of the number of $MoS_2$ layers. The dashed line represents the energy transfer rate to thick, bulk-like $MoS_2$. Blue circles indicate experimental data with associated uncertainty that is smaller than the size of the symbols. The solid line is a fit to Equation 1 (see text for details). Inset: schematic representation of the definition of the thickness (t) and distance (d) parameters.

To quantify the enhancement of energy transfer in our QD/MoS$_2$ system, we compare the energy transfer rate corresponding to the different thicknesses with the energy transfer rate in the bulk limit. The energy transfer rate is given by $k_{ET} = \tau_{DA}^{-1} - \tau_D^{-1}$, and is calculated by spatially averaging data collected from each facet across multiple samples. From a total of five flakes, we obtained data for ten different facets, ranging in thickness from 1 to 8 layers thick. In addition, for visibly thick bulk-like flakes which we estimate to be >20 layers, we obtain an average $k_{ET} = k_{ET}^{3D} = 2.4\ (\pm 0.3) \cdot 10^8$ s$^{-1}$. In Figure 4, we plot $k_{ET}$ as a function of the number of layers, with $k_{ET}^{3D}$ depicted as a dotted line. The enhancement factor $k_{ET}/k_{ET}^{3D}$ (right axis in Figure 4) shows that energy transfer from the quantum dots to a MoS$_2$ monolayer is more than three times faster than energy transfer to the bulk.

Using the continuum model derived by Gordon & Gartstein,[23] we can express the enhancement factor in the case of anisotropic permittivity as a function of the ratio between the flake thickness $t$ and the distance $d$ between the donor and the surface of the acceptor (see inset to Figure 4):

$$\frac{k_{ET}}{k_{ET}^{3D}} = \frac{3}{2\left(\frac{2\pi}{\lambda}t\right)^3} \text{Im}\left[\int_0^\infty d\rho\ \rho^2 e^{-2\rho} \frac{(\beta^2-1)(1-e^{-2\rho\delta t/d})}{(\beta+1)^2 - (\beta-1)^2 e^{-2\rho\delta t/d}}\right], \tag{1}$$

where $\beta = \sqrt{\varepsilon_\parallel \varepsilon_\perp}$ is the effective permittivity of the acceptor, $\delta = \sqrt{\varepsilon_\parallel/\varepsilon_\perp}$ is a parameter representing the dielectric anisotropy, $\lambda$ is the donor emission wavelength, and ρ is a variable of integration. Taking the distance from the center of the QD to the MoS$_2$ surface to be $d = 6.3$ nm and a flake thickness of $t = 0.65$ nm per S-Mo-S layer,[27] we fit the data in Figure 4 to Equation 1, allowing the real and imaginary parts of $\varepsilon_\parallel$ and $\varepsilon_\perp$ to be varied as fit parameters. A least-squares regression yielded the strongly anisotropic dielectric functions $\varepsilon_\parallel$ = 18.2 + 11.3i and $\varepsilon_\perp$ =

7.1, where $\varepsilon_\perp$ has a negligible imaginary component. The large difference in the parallel and perpendicular dielectric functions is consistent with recent reports on the complete in-plane polarization of the direct excitonic transition dipoles in $MoS_2$.[38] Additionally, these values compare favorably to an isotropically averaged dielectric function reported recently for $MoS_2$.[25] It should be noted that the values obtained here represent the "effective" permittivity of $MoS_2$ within our hybrid architecture, since Equation 1 assumes vacuum as the immersing medium whereas our $MoS_2$ flakes are supported between $SiO_2$ ($n \approx 1.5$) and a layer of organic-capped quantum dots ($n \approx 1.8$). Consequently, placing $MoS_2$ on substrates with lower permittivity or suspending over air[1] could further emphasize the enhancement.

In conclusion, we have demonstrated efficient (>95%) energy transfer between zero-dimensional semiconductor quantum dots and a two-dimensional semiconductor TMDC. We have shown that, as a result of reduced dielectric screening in thin layers, the energy transfer rate is enhanced relative to the bulk. Recent theoretical work has suggested that the effective dielectric constant of few-layer TMDCs can be controlled through an externally applied electric field.[39] This would allow for *in-situ* manipulation of the dielectric screening effect and opens up new avenues toward controlled motion of excitonic energy at the nanoscale. These results can lead to improved optoelectronic devices in which the performance is enhanced through surface modification, electrostatic gating, and functionalization of TMDCs.

*Acknowledgements*

We thank J. Kong, X. Ling, A. Castellanos Gomez and E. Navarro Moratalla for discussions. We thank M. S. Strano and S. Shimizu for the use of the Raman microscope, and S. Coe-Sullivan and J. Steckel of QD Vision, Inc., for supplying QD materials. This work was supported as part of the Center for Excitonics, an Energy Frontier Research Center funded by the U.S. Department of Energy, Office of Science, Office of Basic Energy Sciences under Award Number DE-SC0001088 (MIT).